\documentclass[aps,prc,twocolumn,floatfix,nofootinbib]{revtex4}
\usepackage{float,ulem}
\usepackage{graphicx,epsfig,epstopdf,amsmath}

\begin{document}

\title{Jet shape modification in a transport calculation}
\author{Monideepa Maity, Subrata Pal} 
\address{Department of Nuclear and Atomic Physics, Tata Institute of Fundamental Research, 
Homi Bhabha Road, Mumbai 400005, India}

\begin{abstract}
A precise quantification of the medium-modification of the high transverse momenta jets in relativistic
heavy ion collisions rely on consistent modeling of elastic and inelastic energy loss suffered by the 
jet and the concurrent underlying medium evolution. We have developed a unified framework for jet and bulk 
medium evolution within a multiphase transport approach where the jets and medium share energy and 
momentum via multiple elastic scatterings and medium-induced gluon radiation during the parton transport. 
The formulation enables realistic predictions of jet based observables extended to a
large radius of the jet cone in central Pb-Pb collisions at 5.02 TeV.
The model provides reasonable quantitative agreement with the experimental data from inclusive jet suppression 
and the full jet shape function up to large radial distances induced by both the collisional and radiative 
jet energy loss and migration of the lost $p_T$ in the medium.
We find that gradual degradation of jet energy through gluon emissions alters the energy-momentum evolution 
in the jet, essential to describe the entire range of jet shape ratio relative to proton-proton collisions.
Pure collisional energy loss injects appreciable $p_T$ broadening and migration of the medium partons,
resulting in an enhanced population at large angular distances in the jet-shape ratio.

\end{abstract}


\maketitle

\section{Introduction}

Jet quenching has been a key signature in the discovery and characterization of hot
and dense Quark-Gluon Plasma (QGP) formed in heavy-ion collisions
at the Relativistic Heavy Ion Collider (RHIC) \cite{PHENIX:2001hpc,STAR:2002ggv,STAR:2020xiv}
and at the Large Hadron Collider (LHC) 
\cite{CMS:2016qnj,CMS:2018zze,ATLAS:2018gwx,ATLAS:2023hso,ALICE:2019qyj,ALICE:2023waz}.
The high-$p_T$ quark and gluon jets produced in the initial hard collisions suffer about a 
factor of two or more energy loss in traversing the QGP medium as compared to cold nuclear matter. 
Extensive model analysis of the experimental data for the high-$p_T$ single hadron suppression 
and transverse momentum imbalance $A_J = (p_{T,1}-p_{T,2})/(p_{T,1}+p_{T,2})$ between the leading and
subleading jets in a dijet system \cite{ALICE:2013dpt,ATLAS:2010isq}
have established that the quenching in central heavy-ion collisions
is dominated by medium-induced radiative parton energy loss over the collisional energy loss 
\cite{Mehtar-Tani:2013pia,Majumder:2010qh,Cunqueiro:2021wls} which provides 
an important tool in the characterization of the properties of QGP. 

A precise quantification of QGP would require an in-depth microscopic understanding of how the 
medium influence the distribution of the lost energy as well as the jet-induced medium flow and 
medium excitations. 
Jet shapes are commonly employed to describe the $p_T$ distribution of all the 
particles reconstructed within a jet cone of radius $R$ $-$ a distance parameter from the jet axis. 
The full jets, which are sprays of particles produced initially as collimated parton shower radiations 
in vacuum with high virtuality and loses virtuality (become on-shell) 
by medium-induced multi-gluon emission \cite{Majumder:2013re,Cao:2017qpx}, 
provide detailed medium information on modifications of energy-momentum distribution within the jet. 
While the core of the jet comprises hard partons and particles more likely to be produced at the late stages of 
jet evolution, the large radii regime is more sensitive to jet-medium interaction where the
radiated gluons are transported via multiple collisions with the medium partons
(transverse momentum broadening) away from the jet axis. This induces medium
excitations, eventually leading to excess soft particle production \cite{Pal:2003zf}.
Moreover, the diffusion wake generated behind a propagating jet 
\cite{Tachibana:2015qxa,Yang:2021qtl,Yang:2022nei} could lead to depletion of soft hadrons and  
significantly alter the jet transverse profile, i.e. jet cone $p_T$ distribution. 

Early jet shape studies in relativistic heavy ion collisions were mainly confined within
QCD analytic frameworks more suited near the jet axis. Evidently, the modifications of the jet 
substructure and jet shape at large cone radius $R$ depend crucially on realistic heavy-ion 
simulation modeling of the space-time dynamics of medium evolution, the associated collective flow, 
and the complicated interactions between the full jet and the background medium. 
Several theoretical model calculations, 
(MARTINI \cite{Schenke:2009gb}, JEWEL \cite{Zapp:2013vla}, CUJET \cite{Xu:2014ica}, 
Linear Boltzmann Theory (LBT) \cite{He:2015pra,Cao:2016gvr}, jet-hydrodynamic coupled \cite{Tachibana:2017syd})
were employed with varied energy loss formalisms 
(AMY \cite{Jeon:2003gi,Turbide:2005fk}, Higher-Twist \cite{Majumder:2010qh,Cao:2024pxc}, 
DGLV \cite{Gyulassy:2000er,Djordjevic:2003zk}), and
different assumptions of energy-momentum exchange of partons with the background medium. 
Within such coupled jet-hydrodynamics model, studies of jet-induced modifications at radial distances 
$\Delta r = \sqrt{\Delta\eta^2 + \Delta\phi^2}$ far from the jet axis [in relative pseudorapidity 
($\Delta\eta$) and relative azimuth ($\Delta\phi$)] emphasizes the importance of the QGP dynamics 
on jet-shape structures at large jet angles $\theta \gtrsim 1$ \cite{Tachibana:2020mtb}
as well as on the shape of Mach cone induced by hydrodynamic response to the propagating jet 
\cite{Tachibana:2015qxa}, and the accompanying diffusion wakes in the backward region
opposite to the jet direction \cite{Yang:2022nei,Yang:2025dqu}. Within a hybrid weak/strong coupling model, 
characteristic signatures of hydrodynamization of the jet energy was also quantified \cite{Pablos:2019ngg},
and correlations between the jet axis and the broad angular structure of soft particles from 
hydrodynamic freezeout of the diffusion wake was explored \cite{Casalderrey-Solana:2016jvj}.
All these hybrid models combine relativistic hydrodynamic model as background and consider
instantaneous thermalization of the lost energy and momentum into the medium 
and/or assume medium excitations in the form of thermal recoil partons 
\cite{Schenke:2009gb,Xu:2014ica,He:2015pra,Cao:2016gvr}. Whereas, current jet studies 
based solely on parton transport use either simplistic underlying dynamics \cite{Xu:2007aa}
or only collisional energy loss process \cite{Luo:2021hoo}. It is therefore important to perform 
jet quenching studies by developing a concurrent and unified framework (within kinetic approach) 
starting from shower partons formation, jet with medium partons interactions, 
final-state parton hadronization, and involving a reliable underlying partonic-medium dynamics.

In this article, we explore the medium impact on the full jet structure extended to large radial distances 
in a consistent approach within a MultiPhase Transport (AMPT) model \cite{Lin:2004en} that incorporates all the
stages of heavy-ion collisions from parton production and transport to hadron production and transport. 
The current version of the AMPT Monte Carlo event generator incorporates parton kinetic description solely 
via two-body elastic scatterings, has achieved remarkable success in addressing several 
bulk observables pertaining to anisotropic flow and flow fluctuations of hadrons 
\cite{Bhalerao:2014xra}. While collisional energy loss, may to some extent, mimic the jet 
physics at RHIC \cite{Cao:2024pxc,Pal:2003zf}, the predominance of energetic jets and highly 
opaque medium formed at the LHC should enforce radiative gluon showers as the dominant jet energy 
dissipation mechanism and subsequent enhanced interactions of the emitted shower partons with the QGP.
We present the formulation and implementation in the AMPT model of the parton splittings 
to treat the inelastic parton energy loss in collisions involving hard jets and medium partons
and the evolution of the parton showers. With such a complete description of the jet-quenching 
and bulk-parton dynamics in the kinetic framework, we demonstrate the importance of jet-medium interactions 
on how the radiative energy loss and medium response near and at large distances from the jet 
axis provide information on the properties of the QGP medium.

\section{AMPT model with elastic and inelastic parton transport}

The AMPT transport model \cite{Lin:2004en} finds extensive and successful application in studies of
several bulk observables in relativistic heavy-ion collisions, particularly the pair correlations 
and anisotropic flow \cite{Bhalerao:2014xra,Bhalerao:2014mua}, the magnitudes of event-plane correlations 
\cite{Bhalerao:2013ina} and other multiparticle correlations \cite{ALICE:2017kwu}.
The AMPT used in this article employs the HIJING 2.0 model \cite{Deng:2010mv,Pal:2012gf}
to determine the initial nuclear and parton configurations in an event. 
In the two-component HIJING model, nucleon-nucleon collisions with $p_T$ transfer larger than a 
cutoff $p_0$ lead to the production of energetic minijet partons, and the soft interactions 
(with $p_T < p_0$) 
characterized by an effective cross section $\sigma_{\rm soft}$ result in string excitations.
The original AMPT \cite{Lin:2004en} includes the HIJING 1.0 model for initial condition that employs
fixed values for $p_0=2$ GeV, $\sigma_{\rm soft} =57$ mb and the Duke-Owens parton distribution 
function (PDF) for free proton. However, this PDF underestimates the parton densities at small $x$
crucial for minijet production at high energies. Whereas, the updated HIJING 2.0 uses a
center-of-mass energy dependent (and larger) cut-off for $p_0(\sqrt{s_{NN}})$, $\sigma_{\rm soft}(\sqrt{s_{NN}})$ 
and the modern Gl\"uck-Reya-Vogt PDF. For the nuclear PDF, HIJING 2.0 employs a factorized form 
$f_a^A(x,Q^2) = R_a^A(x,Q^2) f_a^N(x,Q^2)$, 
with explicit impact parameter dependent parton shadowing \cite{Deng:2010mv,Pal:2012gf}. 
While HIJING 1.0 model largely overestimates the inclusive jet production cross section $\sigma_{\rm jet}$ 
with increasing energy $\sqrt{s_{NN}} \gtrsim 200$ GeV, the HIJING 2.0 model provides a realistic energy 
dependence of $\sigma_{\rm jet}$ and describes most of the features of hadron production 
in p+p collisions at LHC energies up to 7 TeV \cite{Deng:2010mv}. 
We have considered the string melting version of AMPT where the strings are melted 
into their constituent quarks and antiquarks. The string melting AMPT along with HIJING 2.0 initial conditions
explains remarkably well the centrality dependence of several bulk observables including the anisotropic flow data
at the RHIC and LHC energies \cite{Bhalerao:2014xra,Bhalerao:2014mua,Bhalerao:2013ina,Pal:2012gf}.

Currently, the scatterings and evolution of these (anti-)quarks and minijet partons in AMPT are treated 
in the ZPC parton cascade using only parton-parton elastic scattering cross section. 
In the current study, the AMPT model is developed for jet quenching studies by including 
inelastic collisions via multiple gluon radiation, induced in each collision of the jet with 
the soft medium partons. To ensure high-$p_T$ statistics, in each event, we embed a 
single full-jet triggered from PYTHIA 6.4 \cite{Sjostrand:2014zea} with initial state
radiation and multi-parton interactions. In the final-state high-virtuality parton shower 
post hard scattering, each splitting/formation time can be estimated as 
\begin{equation}
\tau_{{\rm form}, i} = \frac{2 E_i \: z_i (1-z_i)}{k^2_{\perp,i}},
\label{eq:ftime}
\end{equation}
where $E_i$ is the energy of the parent parton and $z_i$ and ($1-z_i$) are the energy fractions carried 
by the outgoing partons with transverse momentum $k_{\perp,i}$ relative to the parent. 
In the multiple splittings scheme \cite{Zhang:2022ctd} employed here, the formation time of 
each parton outgoing parton is taken as $\tau_{\rm form} = \sum_i \tau_{{\rm form}, i}$, obtained 
by summing over the splitting times of all its parent partons starting from the first in the 
full splitting sequence as provided by the PYTHIA shower formation tree. Each of these partons
is then allowed to interact with the AMPT medium partons only at times $t \geq \tau_{\rm form}$. 
Thus the high-virtuality parton shower evolution would gradually develop over time 
\cite{Modarresi-Yazdi:2024vfh}.

The jet shower partons and the soft bulk partons are evolved in Boltzmann transport with elastic and inelastic collisions. 
The collisional energy loss is provided by the 2-parton $ab\to cd$ scattering channels 
which include $gg \to gg$, $gq \to gq$, $qq\to qq$, involving gluons $g$ and quarks $q$ 
and equivalent channels involving antiquarks $\bar q$. The differential elastic scattering 
cross section at the leading order regulated for collinear divergences by the 
Debye mass $\mu$ is given by \cite{Lin:2004en,Pal:2003zf}
\begin{equation}
\frac{d \sigma_{ab \to cd}^{\rm el}}{d\hat t} = C_a 4\pi \alpha_s^2
\left(1 + \frac{\mu^2}{\hat s} \right) \frac{1}{(\hat t - \mu^2)^2} ,
\label{eq:elastic}
\end{equation} 
where $\hat t$ and $\hat s$ are the usual Mandelstam variables, and 
the Casimir factor $C_a = 9/8, 1/2, 2/9$ for $gg, gq, qq$ scatterings. Unless mentioned, throughout our 
calculations we have taken the strong coupling constant $\alpha_s = 0.333$ and the Debye screening mass 
$\mu = 3.226$ fm$^{-1}$ which corresponds to total elastic cross section 
$\sigma^{\rm el} \approx C_a 4\pi\alpha_s^2/\mu^2 = 1.5$ mb for gluon-gluon scatterings.
The string-melting AMPT with these parameter values agrees very well with hadron multiplicity and flow 
harmonics data at LHC energies. It should be noted that energy deposition in the QGP medium from 
elastic scatterings with energetic partons can also lead to jet-induced medium flow and medium excitations 
producing soft hadrons.

Having noted the importance of radiative energy loss channel (i.e. inelastic scatterings) in 
jet quenching studies, we adopt closely the methodology developed for the LBT-hydro model in 
Refs. \cite{He:2015pra,Cao:2016gvr} for the inelastic energy loss for the hard shower partons only.
However, instead of using hydrodynamics as medium, our entire framework including the jet-medium 
interactions is implemented in the kinetic approach. Specifically, each jet-medium interaction in the LBT model
lead to thermal recoil partons scattered out of the hydrodynamic underlying medium which further 
recoil, while the radiated partons excite the medium via elastic and inelastic collisions. 
In contrast, the present kinetic/transport approach includes discrete elastic and inelastic interactions 
between the propagating jet shower partons and the soft medium partons 
formed in AMPT \cite{Lin:2004en,Pal:2003zf} with explicit energy-momentum conservation in each collision as detailed below.    
While these collision processes generally degrades the energy of hard jet, the energy-momentum of the minijet/soft 
interacting partons gets modified which gradually spread via subsequent elastic scatterings with the medium partons
as encoded in the usual AMPT model.

The medium-induced gluon radiation spectrum from the jet partons due to inelastic collisions 
is taken from higher-twist energy loss calculation \cite{Guo:2000nz,Majumder:2009ge} 
\begin{equation}
\frac{dN_g^j}{dz dk_\perp^2 dt} = \frac{2\alpha_s C_A}{\pi} {\hat q}_j
\frac{k_\perp^4 P_j(z)}{(k_\perp^2 + z^2 m_j^2)^4}  \sin^2 \left( \frac{t-t_i}{2\tau_f} \right).
\label{eq:split}
\end{equation} 
Here $k_\perp$ and $z=\omega/E_j$ are the transverse momentum and fractional energy of the emitted gluon
with respect to the parent parton $j$ of mass and energy ($m_j, E_j$) and $C_A = N_c$ is the gluon color factor.
We employ the DGLAP vacuum splitting kernel $P_j(z)$ which represents an incoming 
hard parton $j$ with momenta $k$ to split into two partons carrying momenta $zk$ and $(1-z)k$. 
The divergence at $z\to 0$ is avoided by applying a lower cut off $z_{\rm min} = \mu/E_j$.
A radiated gluon becomes fully developed to participate in further scattering only
after its formation time $\tau_{f,j} = 2E_j z(1-z)/(k_\perp^2 + z^2 m_j^2)$.
Note the formation time $t_i$ of the parent hard parton is updated at each scattering.
Since we model radiative processes as induced by explicit elastic scattering, the jet
transport coefficient ${\hat q}_j = \langle \Delta q_\perp^2\rangle/dt$, defined as the mean 
squared transverse momentum transfer/broadening per unit length of propagation, 
can be computed as \cite{JET:2013cls}
\begin{equation}
{\hat q}_j = \int d^2{\bf q}_\perp \frac{d \sigma^{\rm el}_{ab\to cd}}{d{\bf q}_\perp^2}  \: {\bf q}^2_\perp .
\label{eq:qhat}
\end{equation}

The inelastic scattering rate $\Gamma_{\rm inel}^j$ and the average multiplicity of emitted gluons 
$\langle N_g \rangle = \Gamma_{\rm inel}^j \Delta t$
in the time interval $\Delta t = t - t_i$ can be obtained by integrating out 
Eq. (\ref{eq:split}) but with a weight factor of half for 
$g\to gg$ process to avoid double counting of final states.
This provides the inelastic scattering probability
$P_{\rm inel}^j = 1 - {\rm exp}(-\Gamma_{\rm inel}^j \Delta t)$. 
The outgoing channel for pure elastic collision in AMPT is explicitly determined. 
In case of inelastic process, the number $N_g$ of emitted gluons for the rare events is 
first obtained from a Poisson distribution with the calculated average $\langle N_g \rangle$.
The energy-momentum of each gluon is then sampled from Eq. (\ref{eq:split}) and finally their 
four-momenta is scaled to ensure energy-momentum conservation for the $2\to 2 + N_g$ process
\cite{He:2015pra,Cao:2016gvr}.
These radiated gluons gradually diffuse into the medium and interact with the QGP partons 
only after the formation time $\tau_f$ by inelastic gluon radiations provided these 
are sufficiently energetic with $p_T > p_{T, {\rm cut}} = 2$ GeV and also
via elastic scatterings leading to jet-induced medium excitations.
This choice of the $p_{T, {\rm cut}}$ value distinguishes the low transverse momentum bulk partons 
surrounding the more energetic incoming/produced jet shower partons.
The surviving (final state) jet showers and radiated partons at freeze-out 
hadronize by independent fragmentation using PYTHIA while the soft medium partons
coalesce to form hadrons at freeze-out as incorporated in the original AMPT.

\section{Simulation results on single and full jets}

For phenomenological jet studies, we follow the same jet reconstruction and definition procedures
as used in experimental analysis \cite{CMS:2016qnj,CMS:2018zze}: 
All the final state hadrons in an event are employed with various soft-
and jet-hadron kinematic/acceptance cuts on pseudorapidity $\eta$, azimuthal angle $\phi$ and transverse momenta $p_T$.
Jets are constructed using anti-$k_T$ algorithm encoded in the FastJet framework \cite{Cacciari:2011ma} with a 
chosen jet cone radius $R$ followed by iterative noise/pedestal underlying event subtraction 
in heavy ion collision. Two-dimensional $\Delta\eta-\Delta\phi$ correlations between these hadrons
and the leading/subleading/inclusive jet axes are constructed to obtain signal pair-distributions normalized
by total number of jets $N_{\rm jet}$:
\begin{equation}
S(\Delta\eta,\Delta\phi) = \frac{1}{N_{\rm jet}} \frac{dN^{\rm same}}{d\Delta\eta \: d\Delta\phi},
\end{equation}
where $\Delta\eta \equiv \eta-\eta_{\rm jet}$ and $\Delta\phi \equiv \phi-\phi_{\rm jet}$ are the differences
in the pseudorapidity and azimuthal angle between the hadrons and jet.
Limited acceptance is corrected by constructing mixed-event correlation $M(\Delta\eta, \Delta\phi)$ to obtain 
acceptance-corrected per trigger jet distribution 
\begin{equation}
\frac{1}{N_{\rm jet}} \frac{dN}{d\Delta\eta \: d\Delta\phi} = S(\Delta\eta,\Delta\phi) 
\frac{M(0,0)}{M(\Delta\eta,\Delta\phi)},
\label{eq:corr}
\end{equation}
followed by sideband ($1.5 < |\Delta\eta| < 2.5$) subtraction \cite{CMS:2016qnj} to correct for
the long-range azimuthal correlation associated with collective flow and uncorrelated distributions.
Such a highly nontrivial analysis procedure is crucial to determine the QGP response on the jets 
near and beyond the jet cone radius $R$.
A full three-dimensional hadron distribution $(1/N_{\rm jet})\: dN/(dp_T \: d\Delta\eta \: d\Delta\phi)$ 
can also be constructed by additional binning of the particles in transverse momentum $p_T$ about the jet axis.
This allows to obtain the $p_T$ distribution of hadrons associated with the jets within the jet cone: 
\begin{equation}
\frac{dN}{dp_T} = \frac{1}{N_{\rm jet}} \int d\Delta\eta \int d\Delta\phi 
\frac{dN}{dp_T \: d\Delta\eta \: d\Delta\phi} \Big|_R .
\label{eq:dNdpt}
\end{equation}
In the present study, the model simulations are performed for (0-10)\%  central Pb-Pb collisions 
at the LHC energy of $\sqrt{s_{\rm NN}} = 5.02$ TeV.

\begin{figure}[t]
\centerline{\includegraphics[width=6.0cm]{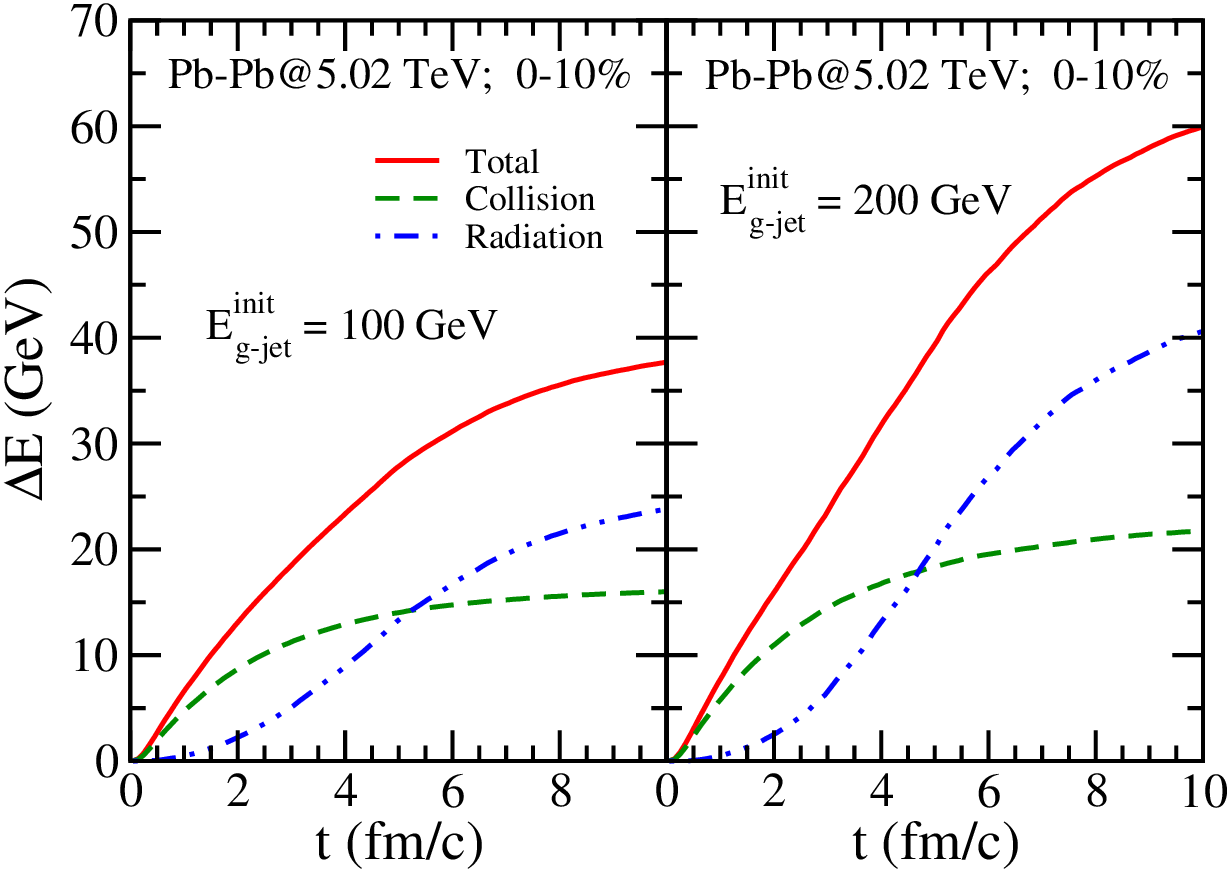}}
\caption{Time dependence of energy loss experienced by an energetic gluon in a jet produced at the center with initial energy 
$E=100$ and 200 GeV in (0-10)\% central Pb-Pb collisions at 5.02 TeV. The results are in the string melting AMPT model for 
elastic scattering (dashed green lines), medium-induced radiation (dash-dotted blue lines) and total (solid red lines).}
\label{fig:Elos}
\end{figure}

Figure \ref{fig:Elos} shows the time dependence of collisional, radiative and total energy loss 
of a single gluon initiating a jet with initial energy $E^{\rm init}_{\rm g-jet} =100$ and 200 GeV 
produced at the center of the medium formed in (0-10)\% central Pb-Pb collisions at 5.02 TeV in 
the extended AMPT model. 
Due to multiple scatterings along its propagation path the degradation 
of jet energy leads to a reduced energy loss at the late stages of evolution and the observed
saturation of $\Delta E$.
(A static QGP medium at temperature $T$ has a logarithmic dependence for the elastic energy loss 
per unit mean-free path $dE/d\lambda \sim T^2 \ln (ET/\mu^2)$ and the transverse momentum broadening 
${\hat q} \sim T^3 \ln (ET/\mu^2)$ in the small-angle scattering approximation \cite{Armesto:2011ht}.)
While the collisional energy loss is seen to be important at $t \lesssim 4$ fm/c, the energy loss from
medium-induced gluon radiation becomes increasingly dominant at later times due to stronger
path-length dependence in the fully formed opaque QGP medium.
For the more energetic gluon $E^{\rm init}_{\rm g-jet} = 200$ GeV, the radiation loss 
is found to be even larger and contributes finally to about 67\% of the total energy loss 
(though its fractional energy loss $\Delta E/E$ is smaller) 
emphasizing its importance in jet quenching studies at LHC energies.

\begin{figure}[t!]
\centerline{\includegraphics[width=6.4cm]{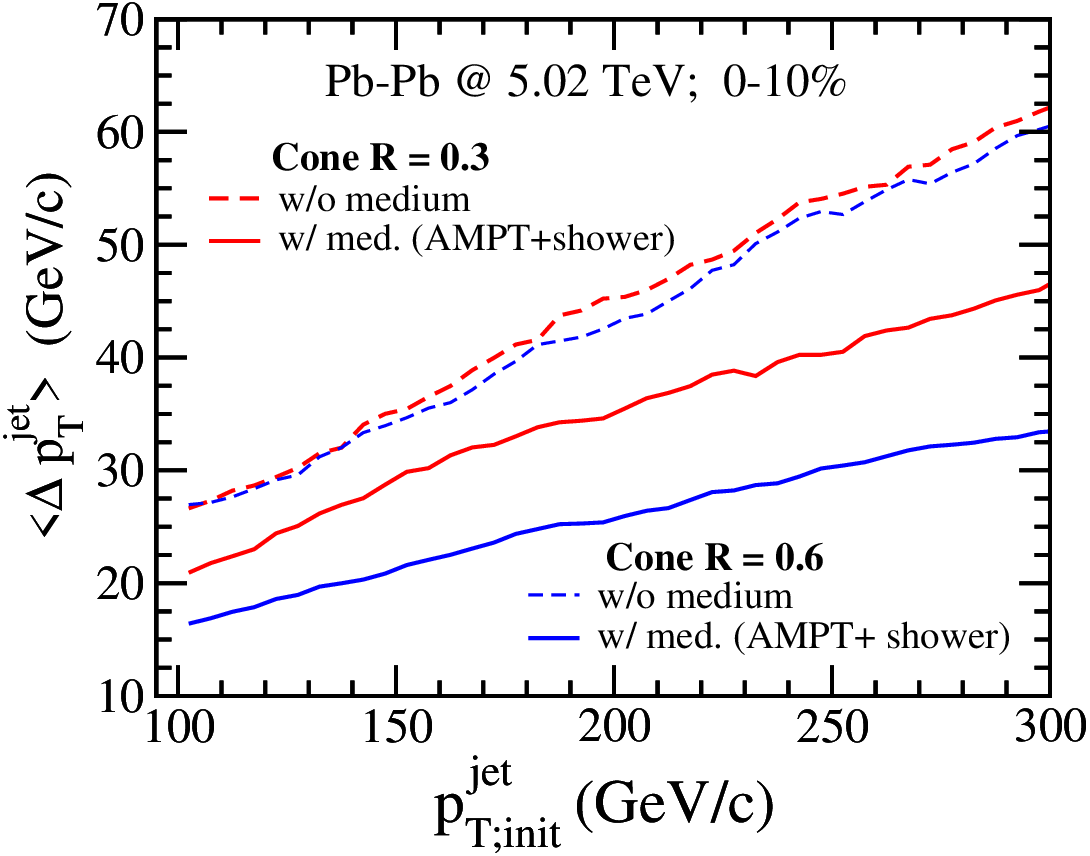}}
\caption{$p_T$ loss of inclusive full jets versus the initial full jet $p_{T, {\rm ini}}^{\rm jet}$ in 
the string melting AMPT model for jet cone sizes $R= 0.3, 0.6$ in (0-10)\% central Pb-Pb collisions at 
5.02 TeV in the absence (dashed lines) and presence (solid lines) of jet-induced medium contribution.}
\label{fig:ptlos}
\end{figure}

In Fig. \ref{fig:ptlos} we present the event averaged transverse momentum
loss of inclusive full jets as a function of the initial full jet $p_T$ reconstructed 
using anti-$k_T$ algorithm with jet-cone sizes of $R = 0.3, \: 0.6$ in (0-10)\% central Pb-Pb collisions.
All the hadrons are included in the jet reconstruction without applying any cuts on the momentum space.
For the shower partons only (without medium response) the effect of cone size $R$ is quite 
small due to collinear radiation near the inner hard core of the jet cone. On inclusion of medium 
contributions derived from interactions of emitted gluons with the QGP partons, the lost 
energy is largely recovered especially for larger cone size $R$. 
This is due to the fact that $p_T$ collisional broadening of the jet-induced partons 
diffuse and propagate to larger angular distances from the jet axis. 
Further, the lost energy is also carried by the recoiled medium partons as a part of jet-induced
medium response/excitation which also contributes to the final jet energy within a given jet cone size.
We remark that in the transport modeling the lost energy-momentum may not thermalize 
instantaneously \cite{Mehtar-Tani:2022zwf} with the medium 
(as adopted in models coupling jet-hydrodynamic evolutions) but can retain some memory of the correlations.
This may have important influence on jet shape population at large angular distances from the jet axis. 
It may be also noted that the steeply-falling jet spectrum with $p_{T,{\rm init}}^{\rm jet}$ causes the 
highly energy jets to be confined within the core producing relatively narrow jets \cite{Brewer:2018mpk}.
However, these energetic jets
suffer enhanced energy loss which propagate to larger distances and produce wider jets.
We find such competing effects between narrowing from falling initial $p_T$ spectra and 
widening from energy loss lead to an overall widening for jets with increasing $p_T$. 
This results in larger recovery of energy loss (i.e. smaller loss $\langle \Delta p_T^{\rm jet}\rangle$) 
with increasing jet cone radius as seen in Fig. \ref{fig:ptlos} for $R=0.6$ (solid blue line).

\begin{figure}[t]
\centerline{\includegraphics[width=6.4cm]{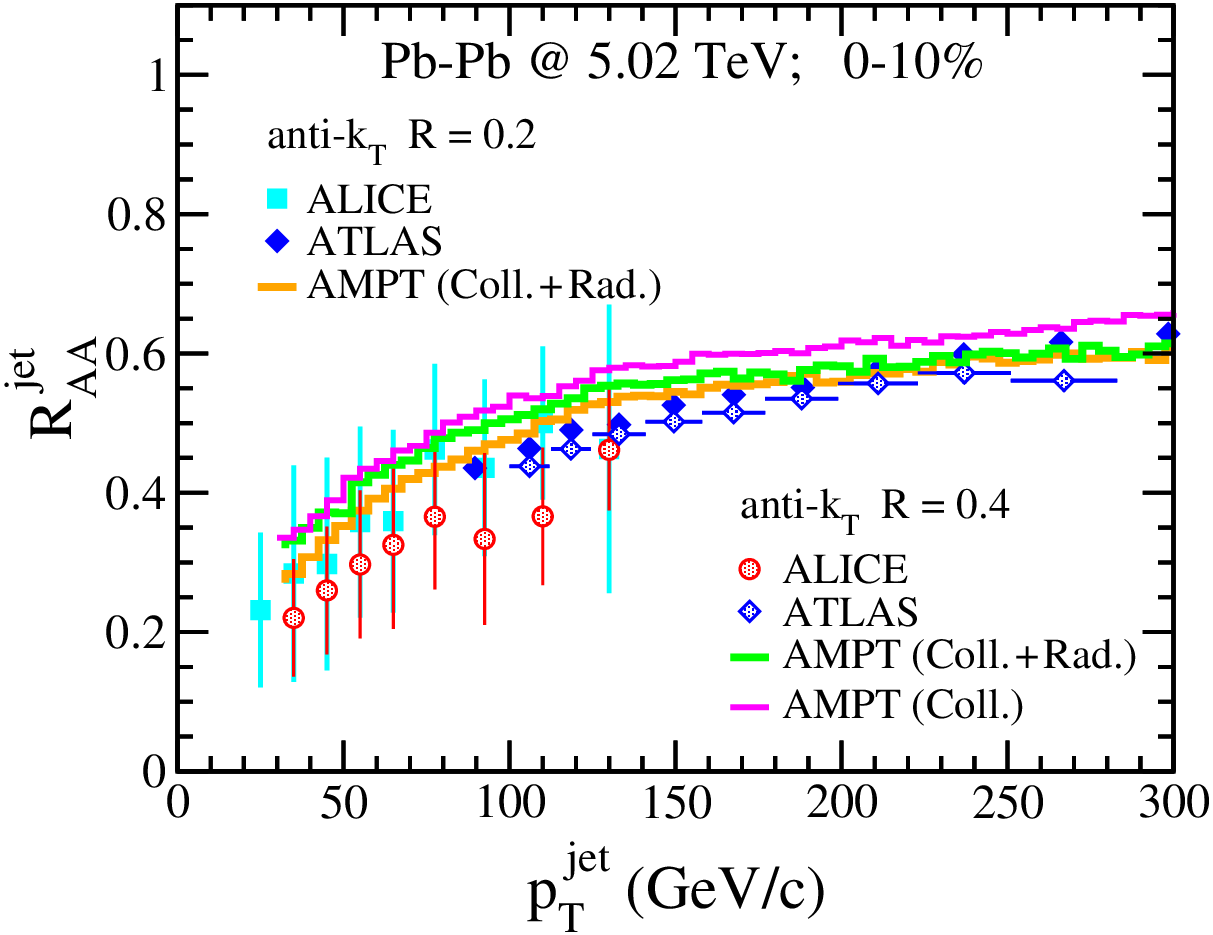}}
\caption{Nuclear modification factor $R_{\rm AA}^{\rm jet}$ for inclusive jets at midrapidity 
reconstructed within cone of radii $R = 0.2, \: 0.4$ in string melting AMPT simulations as compared to data 
from ATLAS \cite{ATLAS:2018gwx,ATLAS:2023hso} and ALICE \cite{ALICE:2019qyj,ALICE:2023waz}
Collaborations in (0-10)\% central Pb-Pb collisions at
$\sqrt{s_{\rm NN}} = 5.02$ TeV. The ATLAS jet has a rapidity coverage  
of $|y_{\rm jet}| \leq 2.0, \:2.8$ for $R=0.2, \: 0.4$ 
while the ALICE $R_{\rm AA}^{\rm jet}$ data includes $|\eta_{\rm jet}| < 0.9 - R$.}
\label{fig:RAAj}
\end{figure}

The full jet energy loss can be quantified by the jet nuclear modification factor 
\begin{equation}
R_{AA}^{\rm jet} = \frac{dN_{\rm PbPb}^{\rm jet}/dp_T d\eta}
{\langle N_{\rm bin} \rangle dN_{pp}^{\rm jet}/dp_T d\eta} ,
\end{equation}
determined by the ratio of single inclusive full jet energy spectrum in Pb-Pb collisions 
relative to baseline p-p collisions and scaled by the average number of binary collisions 
$\langle N_{\rm bin} \rangle$ in Pb-Pb collisions. Figure \ref{fig:RAAj} shows the AMPT results 
of $R_{AA}^{\rm jet}$
as a function of $p_T^{\rm jet}$ for two jet cone radii $R = 0.2$ and 0.4 and confronted 
with experimental measurements by ATLAS \cite{ATLAS:2018gwx,ATLAS:2023hso} and ALICE \cite{ALICE:2023waz}
Collaborations. In this model analysis we also include the hadrons arising from jet-medium interacted 
partons in the jet definition and consider midrapidity jets with $|\eta_{\rm jet}| \leq 2$.
The AMPT calculations with elastic and radiative energy loss
provide reasonable description of the quenching data within the experimental statistical uncertainties.
The model predictions indicate somewhat reduced suppression i.e. larger $R_{AA}^{\rm jet}$ 
for low $p_T^{\rm jet} < 150$ GeV as compared to ALICE data; the under-quenching in the model is seen 
to increase for the larger cone size R=0.4.

Such a model dependence can be explained on the general consideration 
\cite{Casalderrey-Solana:2016jvj,Mehtar-Tani:2024jtd} that low $p_T$ jets contain
predominantly fewer number of (soft) partons where the small amount of transported energy via the 
large-angle scatterings can still be recovered within the larger jet cone. 
With increasing jet $p_T$, the phase-space within the jet 
increases due to more (soft) partons making the jets wider particularly for larger $R$. This results
in larger energy loss or suppression for larger cone size $R$, nearly overcoming the capturing effect
that gives the behavior at low-$p_T$ jets. As a consequence, one observes a smaller sensitivity of 
$R_{AA}^{\rm jet}$ on the jet radius $R$ at large $p_T$. The model prediction of $p_T$ and cone size 
dependence of jet suppression can be used to explore different features of jet-medium interactions, 
namely the jet energy loss and large-angle scatterings. Further, the visible trend in Fig. \ref{fig:RAAj} 
suggests that the under-quenching and opposite $R$-dependence compared to $R_{AA}^{\rm jet}$ data 
may be remedied by injecting larger energy loss for the shower partons which can also overcompensate the
widening due to scattering with increasing $R$. Improvements by including MATTER framework 
\cite{Majumder:2013re,Cao:2017qpx} to simulate in-medium energy loss during the evolution 
of high virtuality hard partons may provide better agreement with the quenching data.

\begin{figure}[t]
\centerline{\includegraphics[width=6.2cm]{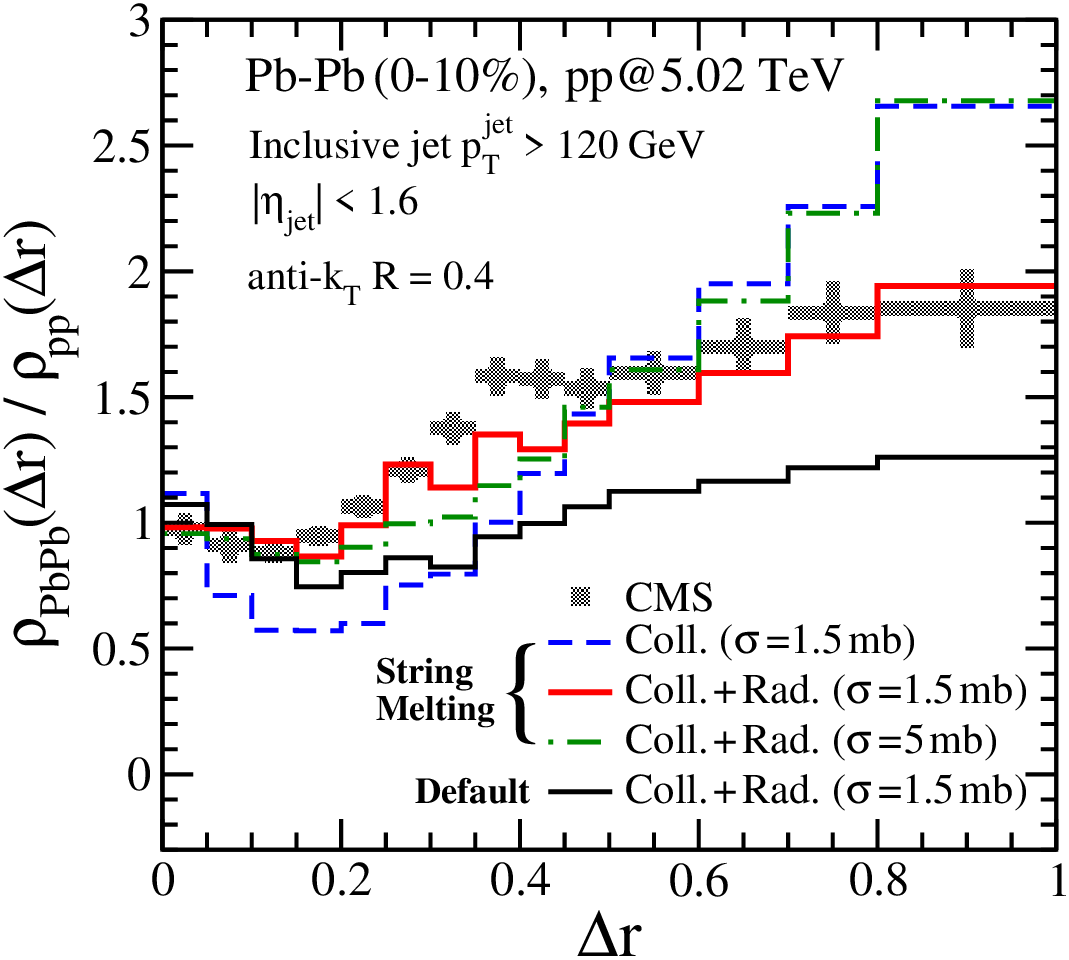}}
\caption{Jet shape ratio $\rho_{\rm PbPb}/\rho_{\rm pp}$ as a function of radial distance $\Delta r$
for inclusive full jets with $p_T^{\rm jet} > 120$ GeV at $|\eta_{\rm jet}| < 1.6$ constructed 
using anti-$k_T$ algorithm with jet cone $R = 0.4$ and cut on hadron $p_T^{\rm trk} > 0.7$ GeV. The CMS data
\cite{CMS:2018zze} for (0-10)\% central Pb-Pb collisions is compared with the string melting AMPT results for 
collisions energy loss only for $gg$ elastic scattering cross section $\sigma^{\rm el}_{gg} = 1.5$ mb 
and collisional plus radiative energy loss for $\sigma^{\rm el}_{gg} = 1.5$ and 5.0 mb. The result from the 
default AMPT model (without string melting) at $\sigma^{\rm el}_{gg} = 1.5$ mb is shown in solid black line.} 
\label{fig:rhorat}
\end{figure}

For orientation we show in Fig. \ref{fig:RAAj} the $R_{AA}^{\rm jet}$ for inclusive jets with only 
collisional energy loss in AMPT. Interestingly, only about 20-30\% contribution from collisional loss 
(as seen in Fig. \ref{fig:Elos}) translates to a relatively large $R_{AA}^{\rm jet}$ suppression 
that is nearly comparable to $R_{AA}^{\rm jet}$ from total energy loss for the low-$p_T$ jets. 
In the absence of radiative energy loss in this case, the initial shower partons are much harder, which
during propagation in the medium can effectively induce large-angle scatterings of the 
soft medium partons and their diffusion. A increasing fraction will be transported out of 
the jet cone leading to an apparent large energy loss 
(see also related discussions in Fig. \ref{fig:rhorat}).

\begin{figure}[t]
\centerline{\includegraphics[width=6.2cm]{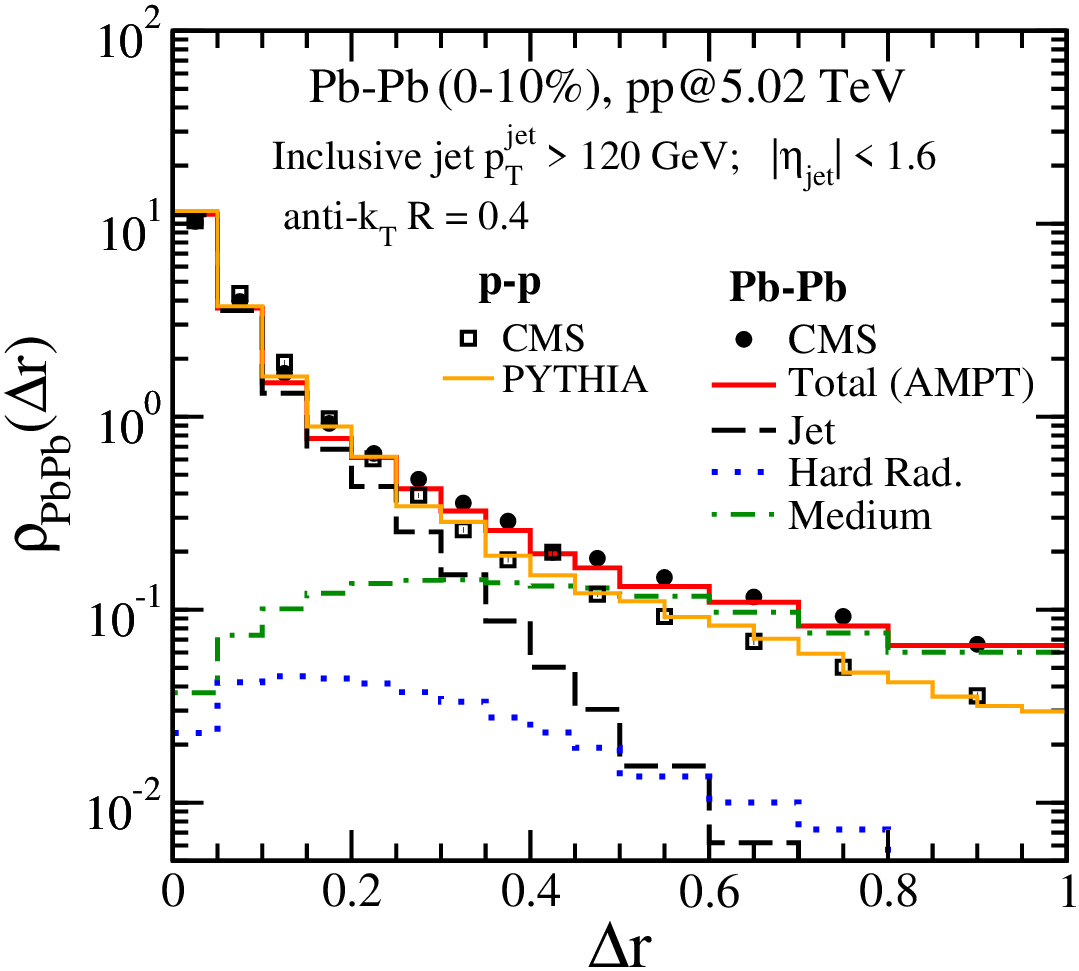}}
\caption{Jet shape $\rho(\Delta r)$ for inclusive full hadron jets with $p_T^{\rm jet} > 120$ GeV at 
$|\eta_{\rm jet}| < 1.6$ constructed using anti-$k_T$ algorithm with jet cone $R = 0.4$ and cut 
on hadron $p_T^{\rm trk} > 0.7$ GeV from the string melting AMPT model (solid red line) compared to CMS 
data (solid circles) \cite{CMS:2018zze} for (0-10)\% central Pb-Pb collisions. The AMPT contributions 
are from fragmentation of final-state hard partons (dashed black line), semi-hard radiated gluons 
(dotted blue line) and jet-induced medium excitations (dash-dotted green line).
Jet shape results from p-p collisions in AMPT/PYTHIA 6.4 (solid orange line) and CMS data (open squares) 
\cite{CMS:2018zze} 
are also shown for comparison.}
\label{fig:rhor}
\end{figure}

To examine the medium modification in detail, we study the jet shape $\rho(\Delta r)$ which describes 
the $p_T$ distribution of all the constituent particles at a radial distance $\Delta r$ from the 
jet axis about a jet cone. One can obtain such a jet transverse momentum profile by constructing 
the acceptance-corrected correlation of Eq. (\ref{eq:corr}) but weighted by the $p_T$ of the 
particles (within a specified range, for instance $p_T \in [0.7, 300]$ GeV) and integrating as:
\begin{align}
P(\Delta r) =& \frac{1}{N_{\rm jet}} \int d\Delta\eta \int d\Delta\phi 
\: \frac{dN}{d\Delta\eta \: d\Delta\phi}  \nonumber \\
& \times \delta(\Delta r - \sqrt{\Delta\eta^2 + \Delta\phi^2}).
\label{eq:prho}
\end{align}
In practice, the jet shapes are constructed as: 
\begin{equation}
\rho(\Delta r) = \frac{N_{\rm norm}}{N_{\rm jet}} 
\frac{ \sum_{\rm jets} \sum_{\Delta r \in [\Delta r_-, \Delta r_+]} p_T^{\rm trk}/p_T^{\rm jet} }
{\Delta r_+ - \Delta r_-} ,
\label{eq:jshape}
\end{equation}
by describing annular rings about the jet axis with 
$\Delta r_\mp = \Delta r \mp \delta r/2$ being the inner/outer annular edges of 
$\Delta r = \sqrt{(\eta_{\rm trk}-\eta_{\rm jet})^2 + (\phi_{\rm trk}-\phi_{\rm jet})^2}$ 
of annulus size $\delta r = 0.05$.
The summation is over all the charged tracks/particles in the rings, and the jet transverse momentum 
profile is normalized to unity within $\Delta r \in [0,1]$. 
Figure \ref{fig:rhor} compares our model prediction of $\rho(\Delta r)$ with CMS data for the full 
inclusive jets of $p_T^{\rm jet} > 120$ GeV at midrapidity $|\eta_{\rm jet}| < 1.6$ constructed with $R=0.4$ 
with radial distance extended to $\Delta r =1$.
The PYTHIA 6.4 result representing baseline proton-proton collision (without any energy loss) agrees remarkably 
well with the jet shape data over the entire range up to $\Delta r =1$, providing the impetus for the model 
study of medium modifications in heavy-ion collisions. 
For central Pb-Pb collisions the jet $p_T$ profile with elastic and radiative channels in AMPT 
(solid red line) shows rapidly decreasing profile up to $\Delta r \approx R =0.4$ beyond which the distribution 
tends to decrease slowly. The qualitative features agree well with the data from CMS Collaborations \cite{CMS:2018zze}.
The shape near the jet core $\rho(\Delta r < 0.2)$ is dictated by the fragmentation/hadronization of the 
collimated leading parton jet and its associated parton showers at the final state that freezes out after 
traversing the QGP medium (dashed black line), with a depletion at larger $\Delta r$ compared to $pp$ collisions. 
While the contribution from semi-hard ($p_T > 2$ GeV) 
radiated gluons (dotted blue line) is found to have a minor effect over the entire shape $\Delta r < 1$, 
the majority of the parton showers are transported by multiple collisional kicks 
(jet-induced flow) to large $\Delta r$ outside the jet cone and can further lead to medium excitations. 
The latter particles are at much lower $p_T$ as these tend to thermalize quite rapidly with the medium, 
and significantly enhance and dominate the jet shape at $\Delta r \gtrsim 0.4$ 
in Pb-Pb collisions (dash-dotted green line).

Figure \ref{fig:rhorat} compares the jet shape ratio $R_{\rm PbPb}^{\rm \rho} = \rho_{\rm PbPb}/ \rho_{\rm pp}$ 
in the AMPT with the CMS data. For pure collisional loss calculations in the current string melting AMPT,
the jet shape ratio (dashed blue histogram) is much smaller for $\Delta r < 0.5$ and more enhanced at large 
radial distances as compared to the data. 
In contrast, the AMPT result with further inclusion of inelastic gluon radiations in 
Fig. \ref{fig:rhorat} (solid red histogram) has an overall better agreement with the experimental data 
especially at large $\Delta r$. This means the medium modifications of the outer core of jet transverse profile 
and at large radial distances are dictated by elastic scatterings. The shower partons (without radiation loss) 
are highly energetic which, during their entire propagation in the QGP medium, can effectively scatter 
the soft medium partons through multiple strong $p_T$ kicks 
to larger distances from the jet axis. Whereas, the degradation of energy of the initial jet partons through collimated
multiple gluon radiation leads to a reduced $p_T$ broadening making the jet shape narrower.
To validate the large-angle collisional-kick mechanism, we re-performed the simulations for radiative 
plus elastic loss with smaller Debye mass $\mu = 1.767$ fm$^{-1}$ (corresponding to $\sigma^{\rm el}_{gg} = 5$ mb) 
which directly increases the strength of elastic scatterings and controls the rate of inelastic splittings
via the jet transport coefficient $\hat q$. While the stronger jet-medium interaction results in somewhat
smaller $R_{AA}^{\rm jet}$ consistent with the suppression data within the error bars (not shown), we find 
a noticeable medium modification of jet shape function [Fig. \ref{fig:rhorat} (dashed-dotted green line)] with 
depletion of particles at small $\Delta r < 0.5$ and an enhanced population of particles at larger distances. 

For orientation, Fig. \ref{fig:rhorat} also compares the jet shape ratio in the default AMPT model \cite{Lin:2004en}.
The evolving medium here consists of relatively much smaller number of minijet partons whereas about 75\% of the 
medium's total energy is associated with the non-interacting soft excited strings (described in Sec. 2).
The significantly reduced parton density entails only fewer radiative and elastic scatterings and exhibits 
a jet shape distinct from the string melting AMPT and data and closer to the proton-proton baseline.
In fact, the success of string melting over the default AMPT model in reproducing the observed large anisotropic flow
for bulk hadrons at RHIC and LHC \cite{Lin:2004en,Bhalerao:2014xra,Bhalerao:2014mua,Bhalerao:2013ina} 
was already realized as due to enhanced parton number and energy densities modeled by melting of excited 
strings into valence (anti)-quarks. We find in this study that jet shapes are also sensitive and could 
potentially delineate the underlying QGP medium density and perhaps their associated evolution.     

It is important to mention that the jet shape ratio for collisional energy loss in the present calculation 
(with $\sigma^{\rm el}_{gg} = 1.5$ mb) shows a distinct behavior from the original string-melting AMPT 
that uses default HIJING 1.0 model \cite{Luo:2021hoo}. Therein, the CMS data is largely overpredicted 
as early as $\Delta r \approx 0.10$, where the enhancement increases rapidly with $\Delta r$
and tends to flatten at $\Delta r \gtrsim 0.50$. This can be traced to 
(i) a smaller and more realistic minijet production in HIJING 2.0 initial conditions 
as discussed in Sec. 2 and (ii) the more dramatic effects due to time-delayed development 
of parton shower during the high-virtuality stage.
Both of these initial state aspects lead to relatively reduced energy-momentum exchange with the 
medium partons and smaller migration of soft particles to large distances.

\section{Conclusions}

We have developed the first unified framework within kinetic approach for parton jet energy 
loss combined with a realistic dynamically evolving QGP medium. This was achieved by employing the AMPT 
model with a proven correct description of the underlying bulk dynamics and made the first important 
implementation of energy loss formalism from medium-induced inelastic splittings in addition to the existing 
collisional energy loss for light quarks and gluons.  
Within this framework, we showed that energy loss is dominated by medium-induced gluon radiations generating 
medium excitations that can be captured for a larger jet cone radius. Higher $p_T$ jets lose larger
energy that diffuses to larger angular distances, producing wider jets and overwhelming the narrowing of jet
profile from steeply-falling initial jet spectra.  
The jet suppression near the low and high $p_T$ jets showed distinct sensitivity on the jet opening angle $R$ 
which creates the possibility to explore different aspects of the jet-medium interactions, namely the dominance 
of scattering at low $p_T$ versus the magnitude of energy loss at high $p_T$. We find that the model 
underpredicts the $R_{AA}^{\rm jet}$  data suggesting a larger energy loss is required especially in the 
high-virtuality stage of parton shower evolution.  

A pure collisional process (without radiative energy loss) may well be characterized  
by the jet shape ratio as a noticeable dip followed by significant enhancement of particle population 
at larger distances ($\Delta r > 0.4$) from the jet axis induced by strong kicks by energetic 
shower partons on the medium partons. However, the $R_{AA}^{\rm jet}$ at low jet $p_T$ appears 
largely indistinguishable from the collisional plus radiative energy loss scenario as these soft 
particles will be naturally missed for small jet cone radius.
Allowing for further gluon shower emission, populates particles about the hard jet core, alters 
medium's energy-momentum spectrum about the jet and provides a good overall description of jet-shape data 
up to large radial distances $\Delta r \in [0,1]$ in Pb-Pb collisions at LHC. 
Another key outcome pertains to sizable differences in the jet shapes 
in the present study of realistic minijet production (in HIJING 2.0) and formation time in the high 
virtuality parton shower as compared to earlier finding in Ref. \cite{Luo:2021hoo}
based on HIJING 1.0 initial condition with unrealistically enhanced (mini)-jet production. 
This suggests a potentially discernible sensitivity of the
jet shape modifications on the initial conditions of parton distribution and their interactions.
Furthermore, while the propensity of large parton density and explosive parton dynamics in reproducing the anisotropic flow 
of bulk hadrons in heavy-ion collisions was well established in the string melting AMPT model,
their significance also in the correct description of jet quenching and jet shape ratio was 
demonstrated in the present analysis.    

The work establishes a crucial step in building a realistic theoretical framework for jet energy loss formalism and 
QGP medium evolution within a non equilibrium transport approach and sets the stage for exploring various energy 
loss formalisms and differential jet substructure observables. This will allow to address more comprehensively the
various aspects of jet quenching in the developed unified framework as compared to other fundamentally 
different yet successful jet-based models.

\vspace{.4cm}

\noindent \textit{\textbf{Acknowledgments:}}
The authors acknowledge financial support by the Department of Atomic Energy (Government of India) under
Project Identification No. RFI 4002.

\end{document}